\documentclass[11pt,article]{iopart}

\usepackage{microtype}
\usepackage{graphicx}
\usepackage{braket}
\usepackage{hypernat}
\usepackage[svgnames]{xcolor}
\usepackage{dsfont}
\usepackage{cite}
\usepackage{iopams}

\expandafter\let\csname equation*\endcsname\relax

\expandafter\let\csname endequation*\endcsname\relax

\usepackage{amsmath}

\newcommand{\ketbra}[2]{\ket{#1}\!\bra{#2}}

\usepackage[separate-uncertainty]{siunitx}

\usepackage{hyperref}
\hypersetup{colorlinks,linkcolor={DarkBlue},citecolor={DarkBlue},urlcolor={DarkBlue}}
\usepackage[capitalize]{cleveref}

\makeatletter
\newcommand\footnoteref[1]{\protected@xdef\@thefnmark{\ref{#1}}\@footnotemark}
\makeatother

\begin{document}
\title[]{Experimental control of the degree of non-classicality via quantum coherence}

\author{A Smirne$^{1,2}$\footnote[3]{\label{f1}These two authors contributed equally}, T Nitsche$^{3}$\footnoteref{f1}, D Egloff$^{1,4}$, S Barkhofen$^3$, S~De$^3$,
I~Dhand$^{1,5,\dag}$, C Silberhorn$^3$,
S F Huelga$^1$, M B Plenio$^1$}
\address{$^1$Institute  of  Theoretical  Physics and IQST,  Universit{\"a}t  Ulm,  Albert-Einstein-Allee~11 D-89069  Ulm,  Germany\\
$^2$Dipartimento di Fisica “Aldo Pontremoli”, Universit{\`a} degli Studi di Milano, e Istituto Nazionale di Fisica Nucleare, Sezione di Milano, Via Celoria 16, I-20133 Milan, Italy\\
$^3$Integrated Quantum Optics Group, Applied Physics, University of Paderborn, D-33098 Paderborn, Germany\\
$^4$Institute  of  Theoretical  Physics,  Technical University Dresden, D-01062 Dresden, Germany\\
$^5$ Present address: Xanadu Quantum Technologies, Toronto, Ontario, Canada
}

\ead{$^\dag$ish.dhand@gmail.com}


\begin{abstract}
The origin of non-classicality in physical systems and its connection to distinctly quantum features such as entanglement and coherence is a central question in quantum physics. 
This work analyses this question theoretically and experimentally, linking quantitatively non-classicality with quantum coherence. 
On the theoretical front, we show when the coherence of an observable is linearly related
to the degree of violation of the Kolmogorov condition, which quantifies the
deviation from any classical (non-invasive) explanation of the multi-time statistics.
Experimentally, we probe this connection between coherence and non-classicality in a time-multiplexed optical quantum walk.
We demonstrate exquisite control of quantum coherence of the walker by varying the degree of coherent superposition effected by the coin, and we show a concomitant variation in the degree of non-classicality of the walker statistics,
which can be accessed directly by virtue
of the unprecedented control on the measurement-induced effects obtained via fast programmable
electro-optic modulators.
\end{abstract}

\maketitle
\section*{Introduction}%
Which predictions of quantum mechanics can and which cannot be reproduced by means of any plausible classical theory? 
This question is at the foundation of upcoming quantum technologies including sensing, computation and communication.
At a more fundamental level, the question is central to determine if certain phenomena are genuinely quantum, for instance in biological or thermodynamical systems~\cite{Engel2007,Huelga2013,Plastina2014,Gonzalez2019,Deffner2019,Wang2019}.

Different strategies have been developed to assess the quantumness of physical systems without having to rely on the knowledge of the microscopic details of the system at hand.
These strategies rely, instead, on directly evaluating the probability distributions of the measurement outcomes with respect to specific traits of classical statistics, such as locality~\cite{Bell1987}, non-contextuality~\cite{Kochen1967,Cabello2014}, and measurement non-invasiveness~\cite{Leggett2002}.
In particular, the latter means that one can access, at least in principle, the value of an observable without altering the statistics associated with its sequential measurements at different times.
Non-invasiveness is indeed strictly related to the Leggett-Garg inequalities~\cite{Leggett1985,Emary2014,Robens2015,Zhou2015}; as well as to the notion of non-signalling-in-time~\cite{Clemente2016,Knee2016,Halliwell2017}; and, ultimately, to the very defining property of classical stochastic processes, i.e., the validity of the Kolmogorov (consistency) conditions~\cite{Feller1971,Milz2017}.

Moreover, the notion of non-classicality of multi-time statistics as specified above is intimately connected with a key resource of quantum systems, namely quantum coherence~\cite{Baumgratz2014,Streltsov2017}.
Specifically, Ref.~\cite{Smirne2018} shows that, under precisely-defined circumstances, the statistics obtained from 
sequential measurements at different times cannot be traced back to classical statistics as defined by the Kolmogorov conditions~\cite{Feller1971}, if and only if coherences are first generated and subsequently turned into populations
in the course of the evolution.
The current work further analyses this connection theoretically and demonstrates its validity in a photonic quantum-walk experiment.

Quantum walks represent a well-established framework to investigate to what extent we can detect and control intrinsically quantum behaviours~\cite{Aharonov1993,Knight2003,Kempe2003,Du2003,Perets2008,Schmitz2009,Karski2009,Childs2009,Sansoni2012,Poulios2014,Gualtieri2019}.
An especially promising platform is that of time-multiplexed optical quantum walks~\cite{Schreiber2010,Schreiber2011,Nitsche2016,Schreiber2012,Elster2015}, wherein the position degree of freedom of the walker is encoded into the time domain and the coin degree of freedom is encoded in the polarisation of light.
Such a platform enables controlling the couplings between different positions and therefore of the coherences present in the setup.
Moreover, coherences are conserved for many steps of the dynamics because of the low experimental de-phasing values afforded by the stable optical feedback loops comprising the setup.
Finally, the possibility to out-couple deterministically the optical signal in the course of the evolution via fast electro-optic modulators addressing individual positions of the walker allows one to probe measurement-induced effects into the statistics of the walk~\cite{Nitsche2018}.

\begin{figure}
\centering 
\includegraphics[width=0.7\columnwidth]{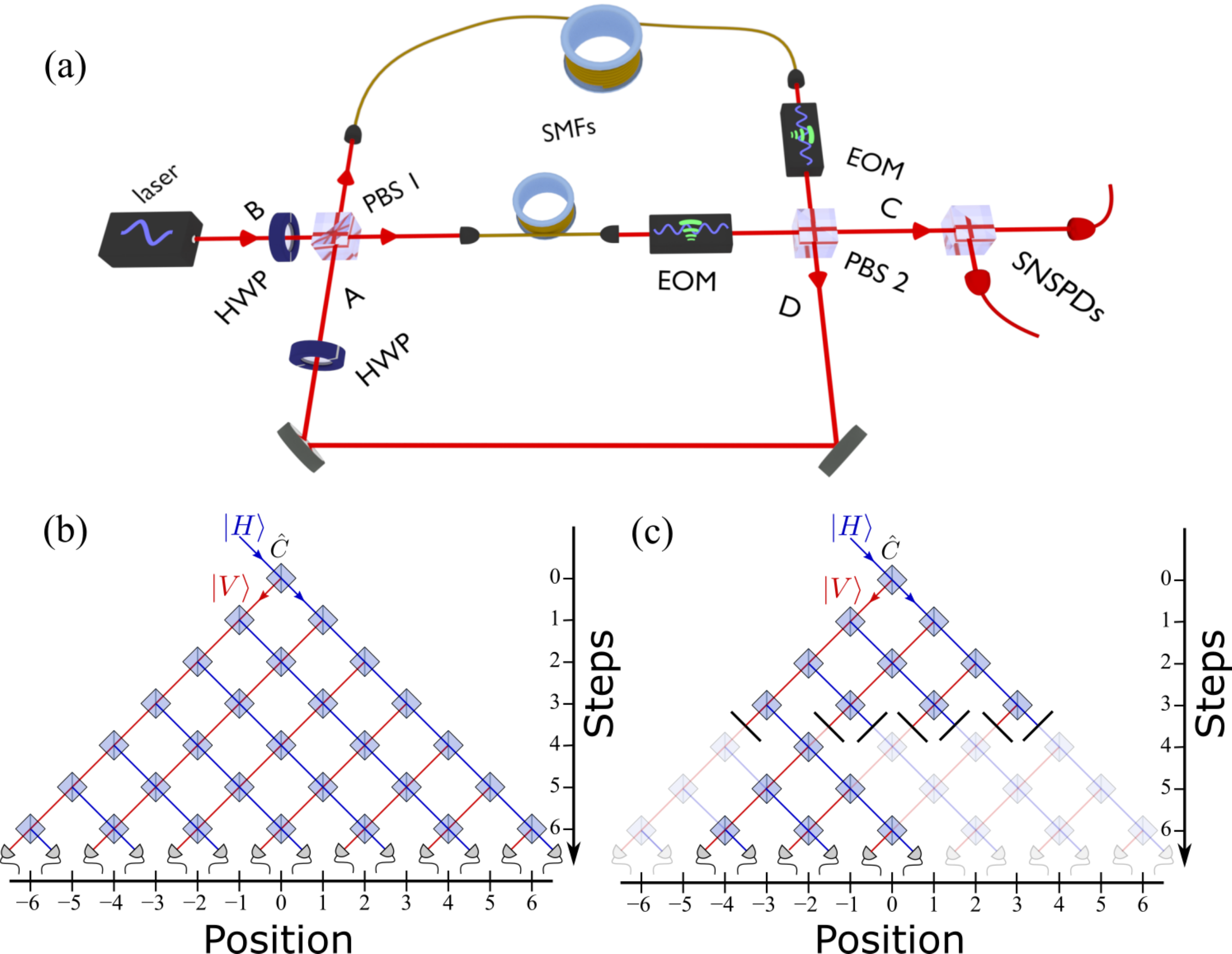}
\caption{\textbf{(a) Schematic of the implemented setup.}
See text for working principle and meaning of acronyms.
\textbf{Quantum walk configurations without (b) and with (c) intermediate measurements.}
(b): Evolution over $N$ steps, without any intermediate measurement.
(c): Evolution in which all  but one mode are out-coupled after step $M$  to perform an intermediate measurement.
$N=6$ and $M=3$ in these sub-figures.
The short black bars in (c) depict out-coupling of the light.
The beam splitter symbols depict the coin toss in addition to the usual PBS operation.
The blue and red lines denote the horizontally and vertically polarised light, which are shifted to the right and the left respectively.}
\label{fig:setup}
\end{figure}

In this work, we study theoretically and experimentally the relation between non-classicality and quantum coherence.
Theoretically, first we make quantitative the connection between quantum coherence and non-classicality derived in Ref.~\cite{Smirne2018}.
Specifically, we show that the violation of the Kolmogorov conditions is directly proportional to the amount of quantum coherence (of the measured observable) that is first generated and later detected.
We then verify experimentally such a relation in a time-multiplexing quantum walk using the setup depicted in \cref{fig:setup}.
Non-classicality is measured by performing sequential measurements of position and coin.
The evolution of quantum coherence in the setup is tuned by controlling the coupling between walker positions.
By changing the coherences, we can tune the violation of the Kolmogorov conditions, thus witnessing a controllable impact of measurement invasiveness and the departure from any classical description of the walk.
We demonstrate unprecedented control and intermediate measurements in a multi-step quantum walk, in this way fully appreciating the non-trivial behaviour of quantum coherence and its effects on non-classicality.

\section*{Quantitative connection between non-classicality and quantum coherence}\label{sect:res}%
Experimentally, we focus on a discrete-time quantum walk on a line, associated with the Hilbert space spanned by the states 
$\left\{\ket{x, c}\equiv \ket{x}\otimes \ket{c}\right\}_{x\in \mathds{Z}; c=H,V},$
where $x$ denotes the position of the walker and $c$ the value of the coin, acknowledging already its experimental realisation in horizontal ($H$) and vertical ($V$) polarisation.
Henceforth, we use the term coin and polarization interchangeably depending on the context.
The initial state is taken to be of the form 
\begin{equation}\label{eq:init}
\rho_0 = \ketbra{x_0}{x_0} \otimes (p \ketbra{H}{H} + (1-p)\ketbra{V}{V}),
\end{equation}
and the evolution is fixed by a unitary operator $\hat{U}$ acting on both the position and coin degree of freedom, so that the global state after $N$ steps is $\rho_N=\mathcal{U}^N \rho_0$, with the unitary super-operator \cite{Rivas2012} $\mathcal{U} \rho = \hat{U} \rho_0 \hat{U}^{\dag}$.
Although our experimental realization is based on quantum walks, the explicit form of the unitary need not be specified for the theoretical treatment presented here.
Even more, the following analysis is valid for a broad class of quantum processes and initial states as discussed in \ref{App:Relation}.

The non-classicality of quantum processes can be assessed unambiguously from sequential measurements of the same observable at distinct times~\cite{Smirne2018,Milz2019,Strasberg2019,Diaz2020}.
These tests of non-classicality compare the statistics obtained from one-shot projective measurements at different final times and the statistics involving projective measurements at intermediate times, thus witnessing the unavoidably invasive nature of measurements in the quantum domain.
Specifically, let $P_{x_0,p}(x,c,N)$ be the probability of having the position $x$ and the coin in $c$ after $N$ steps, given initially the position $x_0$ and the coin value $H$ with probability $p$ as in \cref{eq:init}.
Moreover, let $P_{x_0,p}(x,c,N | y,c',M)$ be the probability of the walker being at position $x$ and the coin in $c$ after $N$ steps, but now conditioned on the fact that after $M$ steps the walker was in position $y$ and the coin in $c'$ (once again, given the initial state fixed by $x_0$ and $p$).
The Kolmogorov conditions~\cite{Feller1971} imply that whenever the statistics of the sequential measurements can be described via a classical stochastic process, the quantity 
\begin{equation}\label{eq:K}
{\sf{K}}_{x_0,p} = \sum_{x,c} \Big |  \sum_{y,c'} 
P_{x_0,p}(x,c,N | y, c',M) P_{x_0,p}(y,c',M)   - P_{x_0,p}(x,c,N) \Big|
\end{equation}
is equal to 0 for any $x_0$ and $p$.
Conversely, any value of ${\sf{K}}_{x_0,p} \neq 0$ signifies the invasiveness of the measurement performed at the intermediate time $M$.
In particular, a non-zero value would exclude any classical description of the walk,
given in terms of the walker possessing definite (even if unknown) position and coin values at all times, which are accessed by the ideal  projective measurements without altering the subsequent walk.


First, we show that it is possible to relate quantitatively the degree of violation of the Kolmogorov conditions, as quantified via ${\sf{K}}_{x_0,p}$, defined in \cref{eq:K}, to the amount of coherences of the measured observable (i.e., values of the off-diagonal elements of the density matrix expressed in the basis of the measured operator) that are generated by the evolution and subsequently turned into populations (i.e., the diagonal elements of the density matrix).
It is this connection that opens the possibility of experimentally controlling the degree of non-classicality by tuning the coherences in the system, as described below.

Consider the quantity~\cite{Smirne2018} 
\begin{align}\label{eq:C0}
{\sf{C}}_{x_0,p} = \left\|\left(\Delta \circ \mathcal{U}^{N-M} \circ \Delta \circ \mathcal{U}^{M} 
- \Delta \circ\mathcal{U}^{N}\right)\rho_0\right\|_1,
\end{align}
where $\circ$ denotes the composition of maps, $\Delta= \sum_{x,c}\ketbra{x,c}{x,c} \cdot \ketbra{x,c}{x,c}$ is the total dephasing map with respect to the measured observable, which in our case is associated with the joint values of the position and the coin, and $\| \cdot\|_1$ is the trace norm.
The measure ${\sf{C}}_{x_0,p}$ quantifies the coherences generated and detected 
by the dynamics.
More precisely, ${\sf{C}}_{x_0,p}=0$ if and only if, starting from the state $\rho_0$ in \cref{eq:init}, no coherences can be generated during the first $M$ time steps and detected with a measurement after $N$ steps of the dynamics.
This notion is strongly connected to important concepts in coherence theory, including the maximal set of incoherent operations and the coherence non-activating set~\cite{AAberg2014,Liu2017},
as discussed in \cite{Smirne2018}.
In addition, we note that the quantifier of coherence defined by Eq.(\ref{eq:C0}) is strictly related to the ``quantum witness" $\mathcal{W}_Q$ introduced in~\cite{Li2012}.
In fact, for the discrete-time unitary evolutions, ${\sf{C}}_{x_0,p}$ reduces to
$\mathcal{W}_Q$ when the total dephasing map $\Delta$ after $N$ steps is
replaced with the projector into one specific population of the reference observable.
In~\cite{Li2012}, where general open-system evolutions and possibly degenerate observables were taken
into account, $\mathcal{W}_Q$ was shown to witness that the
global state at the intermediate time cannot be expressed
as a separable mixture of  system-environment states without coherent components.

As said, we focus on unitary evolution, which is reasonable for our quantum-walk experiment because of its low dephasing rates.
The restriction of unitarity is however not strictly required and we show in \ref{App:Relation} that the same relation between ${\sf{K}}_{x_0,p}$ and ${\sf{C}}_{x_0,p}$ can be derived 
under more general assumptions (namely, that the dynamics is described by a Lindblad equation \cite{Breuer2002} and
that the quantum regression theorem~\cite{Lax1968,Swain1981,Breuer2002,Guarnieri2014,Li2018} holds).
The quantifier defined in \cref{eq:C0} can be expressed with respect to one-time probability distributions, according to
\begin{equation}\label{eq:C}
{\sf{C}}_{x_0,p} = 
 \sum_{x,c} \Big|  \sum_{y,c'} 
P_{y,c'}(x,c,N-M) P_{x_0,p}(y,c',M)  - P_{x_0,p}(x,c,N) \Big|
\end{equation}
(where, with a slight abuse of notation, we denote as $P_{x,c}$ the probabilities when the initial state is $\ket{x,c}$).
Using once again the unitarity, we can then see that 
\begin{equation}
\label{eq:ck} {\sf{K}}_{x_0,p} ={\sf{C}}_{x_0,p}.
\end{equation} 
Eq.(\ref{eq:ck}) is the main theoretical finding of this work; this result will be verified experimentally in the following of the paper.
Crucially, Eq.(\ref{eq:ck}) tells us that controlling the amount of coherences which are generated and turned into populations is equivalent to controlling the degree of non-classicality of the quantum walk.
Such a conclusion can be drawn relying directly on the probability distributions associated with one-time and two-time sequential measurements of the relevant observable; e.g., no full state or process tomography with respect to the position and coin state is needed, which would be certainly a challenging task in most platforms.

We also note that \cref{eq:ck} extends significantly the one-to-one correspondence between non-classicality and the capability of the dynamics to generate and detect quantum coherence shown in \cite{Smirne2018}; such a one-to-one correspondence in fact only means, in terms of the quantifiers introduced above, that ${\sf{K}}_{x_0,p} \neq 0$ if and only if ${\sf{C}}_{x_0,p} \neq 0$, but it does not yield a quantitative connection between ${\sf{K}}_{x_0,p}$ and ${\sf{C}}_{x_0,p}$ when they are non-zero.

\section*{Experimental setup}%
Here we present the salient features of the quantum-walk experiment performed to investigate the correspondence between non-classicality and coherence, and especially the relation \cref{eq:ck}.
Each step of a quantum walk comprises two operations $\hat{U}=\hat{S}\hat{C}$, where the coin flip $\hat{C} = \mathds{1}_x \otimes 
\begin{pmatrix}
\cos(\theta) & \sin(\theta)\\
\sin(\theta) & -\cos(\theta)
\end{pmatrix}$ acts only on the coin degree of freedom, while the conditional shift operator 
$\hat{S} = \sum_x \left( \ket{x + 1}\!\bra{x}\otimes \ket{H}\!\bra{H} + \ket{x - 1}\!\bra{x}\otimes \ket{V}\!\bra{V}\right)$
moves the walker on the line to the right (left) when its internal coin state is $\ket{H}$ ($\ket{V}$). 
These operations are realised with a well-established time-multiplexing architecture~\cite{Schreiber2010,Schreiber2011,Nitsche2016,Schreiber2012,Elster2015} based on an unbalanced Mach-Zehnder interferometer as shown in \cref{fig:setup}(a).

The walker is implemented using a coherent laser pulse (depicted as arriving from the left in \cref{fig:setup}) at a wavelength of \SI{1550}{\nano\meter} with adjustable initial polarisation as the coin degree of freedom.
Such states of light can be used to implement single-particle quantum walks because of the equivalence between single photons and coherent states under quantum walk evolutions (See the \ref{Sec:AppendixEquivalence} and Ref.~\cite{Paul2004}).

The experiment proceeds as follow.
First, a polarisation-dependent splitting is carried out by a polarising beam splitter (PBS) denoted PBS 1 in \cref{fig:setup}(a).
Subsequently, single-mode fibres (SMFs) translate the walker position into the temporal domain by introducing different delays in the two arms.
The setup is closed by an optical feedback loop, thus implementing the conditional shift operator $\hat{S}$ defined above.
Dynamic switches implemented via electro-optic modulators (EOMs) route the pulses either to the detection or back into the feedback loop.
Thus, they capacitate control over whether the dynamics is continued or interrupted and enable the intermediate measurements.
Pulses continuing in the feedback loop will be subjected to the coin operation $\hat{C}$ implemented by a half-wave plate (HWP) before the polarisation-dependent split is repeated, i.e.
they continue the evolution governed by the unitary $\hat{U}$.
On the other hand, the light emitted out of the feedback loop (from port C) is led to the polarisation-resolving detection unit comprising another PBS and two superconducting nanowire single-photon detectors (SNSPDs) where the evolution ends.

The fast EOMs can address individually the different positions values within the walk and thus enable implementing position- and coin-dependent out-coupling.
Such out-coupling corresponds to position-dependent losses, which can be harnessed to perform measurements at intermediate steps of the evolution.
\cref{fig:setup}(b) depicts the unperturbed evolution of the walker over $N$ steps. 
At the end of the evolution, all the light is coupled out and measured.
The corresponding intensity profiles provide the probability distribution $P_{x_0,p}(x,c,N)$, for different values of the final position $x$ and polarisation $c$.
In contrast, \cref{fig:setup}(c) depicts out-coupling of all but one mode, i.e., one polarisation at a specific position, in an intermediate step $M$ of the evolution.
For any walker in this chosen mode, the walk is continued up to step $N$ and the intensity distribution measured.
Any photon detected after $N$ steps would have occupied the chosen mode at step $M$, which means that the selective out-coupling corresponds to a projective measurement of the position and coin of the walk.
This allows us to register the probabilities $P_{x_0,p}(x,c,N | y, c',M)$ of a walk over $N$ steps, with an intermediate measurement at step $M$.
We note that only light intensity measurements, rather than correlation measurements, are required to be performed at the end of the walk to obtain the desired probabilities.
Crucially, the full statistics involving intermediate measurements can be obtained by virtue of an extension of the experiment presented in \cite{Nitsche2018}, harnessing the full flexibility of the setup by coupling out arbitrary positions with specific polarization. This requires a precise control of the extinction ratio of the EOMs in use, which can only be achieved after modifying the Pockels cells in use to suppress piezooptic effects.

By altering the angle of the HWP fixing the coin operation $\hat{C}$, different quantum walks can be realised.
In this way, we can control the amount of coherence generated and detected, and consequently the degree of violation of the Kolmogorov condition according to \cref{eq:ck}.

\begin{figure}
	\includegraphics[width=0.99\textwidth]{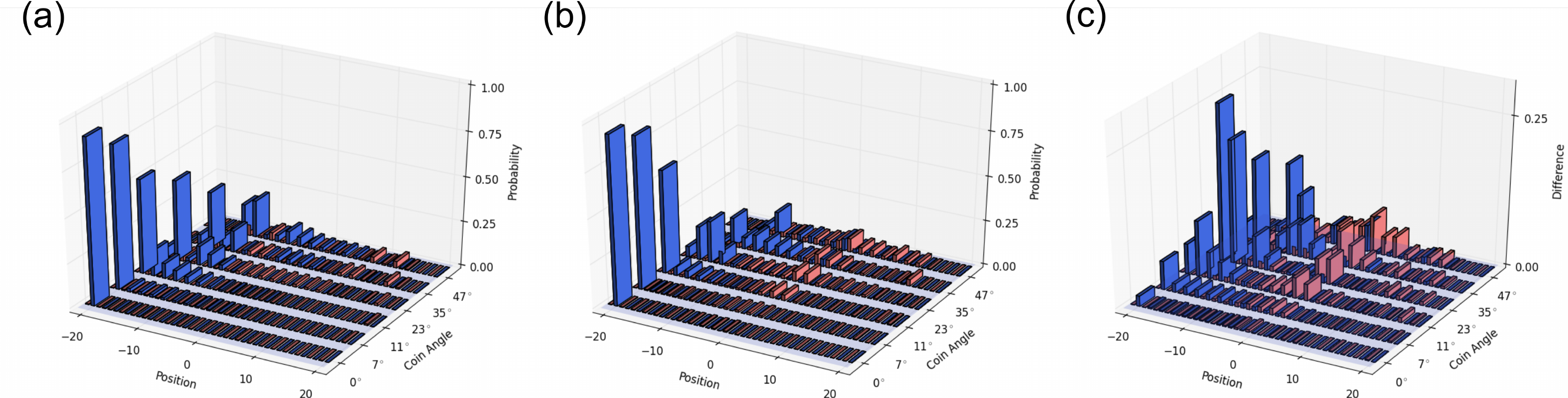}
	\caption{
	\textbf{Visualisation of the quantum coherence generated and detected by the dynamics.} 
	(a): Probability distribution $P_{0,V}(x,c,N)$ as a function of the position $x$ and polarisation $c$.
	(b): Combined probability distribution $\sum_{y,c'} P_{y,c'}(x,c,N/2) P_{0,V}(y,c',N/2)$ as a function of $x$ and $c$. 
	(c): Difference between the two distributions. 
	This difference signals the generation of coherences in the first $N/2$ steps of the walk and their conversion to population after $N$ steps, as presented in \cref{eq:C} 
	For all panels, the initial polarisation is $V$, the initial position $0$ and $N=20$; 
	different rows depict different coin angles $\theta$;
	different columns correspond to different positions $x$ with coin value $V$ in blue and $H$ in red.
 	}
	\label{Fig:Visualize}
\end{figure}

\begin{figure}
\centering
	\includegraphics[width=0.65\columnwidth]{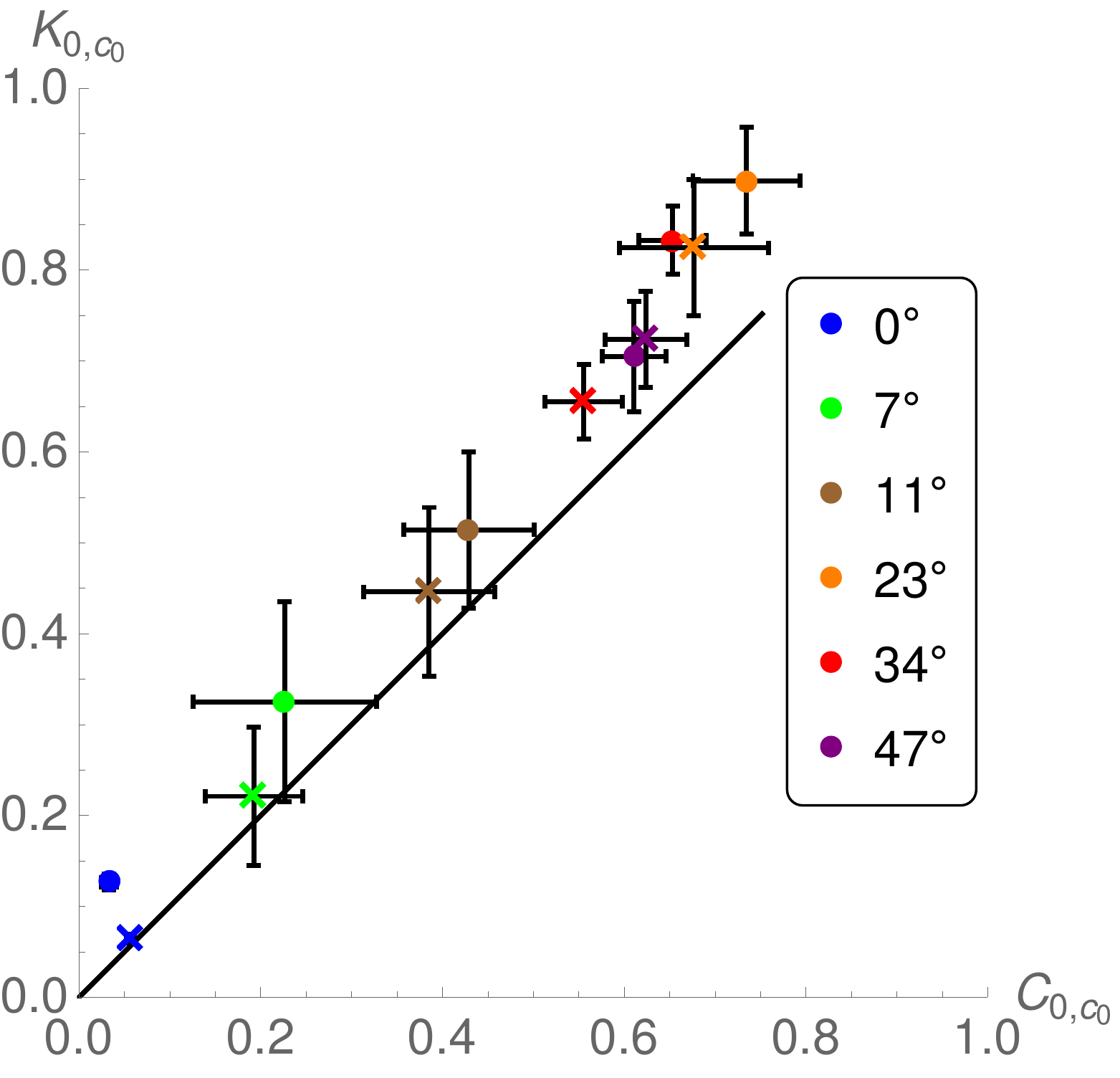}
	\caption{
	\textbf{Linear relation between coherence and non-classicality.}
	We plot on the abscissa the quantifier ${\sf{C}}_{0,c_0}$ of coherences generated and detected 
	as defined in \cref{eq:C} and on the ordinate the degree ${\sf{K}}_{0,c_0}$ of nonclassicality 
	as defined in \cref{eq:K}, for initial polarisation $c_0=H$ (circles) and $c_0=V$ (crosses).
	The black line represents the expected proportionality relation as in \cref{eq:ck}.
	Different points refer to different coin angles.
	Error bars are generated from a Monte-Carlo approach detailed in \ref{App:Errors}.
	A systematic deviation observed in the higher-than-expected ${\sf{K}}_{0,c_0}$ values can be explained by imperfect intermediate measurements as described in the \ref{App:KDisplacement}.
	}
	\label{Fig:KCPlot}
\end{figure} 

\section*{Experimental results}%
We report now the quantum coherences that are generated and detected at different steps of our time-multiplexed quantum walk experiment, and how these are unequivocally related to the degree of non-classicality of the walk itself.
We consider three configurations.
The first two comprise the standard quantum walk of, respectively, $N = 20$ and $M=N/2 = 10$ steps, with the preparation of an initial state $\rho_0$ as in \cref{eq:init} and the measurement of both the position and the coin at the end of the walk as depicted in \cref{fig:setup}(b).
This measurement corresponds to recording the intensity of light for different positions and polarisations.
Results from these two configurations are enough to reconstruct the coherence quantifier ${\sf{C}}_{x_0,p}$ of \cref{eq:C}. 
On the other hand, to evaluate the non-classicality quantifier ${\sf{K}}_{x_0,p}$ defined in \cref{eq:K}, we realize a third quantum-walk configuration, involving both an intermediate measurement after $M=10$ steps and the final measurement after $N = 20$ steps as depicted in \cref{fig:setup}(c). 
We take the initial position $x_0=0$ and the initial pure horizontal or vertical polarisation states; the statistics corresponding to other initial conditions as in \cref{eq:init} can be obtained using the spatial translational invariance of the setup and mixing the probabilities related to pure polarisation states. 

In \cref{Fig:Visualize}(a), we present the probability distribution $P_{0,V}(x,c,N)$ obtained from one-shot position and coin measurements after $N=20$ steps. 
\cref{Fig:Visualize}(b) depicts the quantity $\sum_{y,c'} P_{y,c'}(x,c,N/2) P_{0,V}(y,c',N/2)$, which is obtained by combining one-shot probability distributions obtained from different outcomes of position and coin measurements
after $M=N/2$ steps. 
The difference between the two distributions, reported in \cref{Fig:Visualize}(c), provides us with a clear visualisation of the amount of position and coin coherences which are generated and subsequently turned into populations in the course of the quantum walk, see \cref{eq:C0,eq:C}. 
Indeed, the value of the coin angle determines the coupling between different positions in the walk via the coin degree of freedom, in this way influencing the generation of coherences and consequently their transformation to populations. The different rows of \cref{Fig:Visualize} show how a more balanced coin leads to more coherences and therefore to larger differences in the plotted quantities.

Most importantly, the two-fold interconversion between populations and coherences 
can be related quantitatively to the non-classicality of the quantum walk, according to \cref{eq:ck}.
In order to verify experimentally this relation, we measure the probability distributions $P_{0,c_0}(x,c,N | y, c',M)$ (both for initial polarisation $c_0=H$ and $c_0=V$) associated with the third configuration, which involves an intermediate measurement of the position and the polarisation after $M=10$ steps and a final measurement after $N=20$ steps. 
Combining these probabilities with those obtained from the previous one-shot measurement runs, we can obtain the degree of violation of the Kolmogorov condition ${\sf{K}}_{0,c_0}$ as defined in \cref{eq:K}, for different values of the coin angles and initial coin values. 
The results are reported in \cref{Fig:KCPlot}, where ${\sf{K}}_{0,c_0}$ is plotted against the corresponding value of the generated and detected coherences ${\sf{C}}_{0,c_0}$. 
The equivalence expressed in \cref{eq:ck} is well confirmed by the experimental data, within the error bars. 
This proves a correspondence between the evolution of coherences and the non-classicality in our setup, which is not only qualitative, but strictly quantitative.
As a consequence, we can state unambiguously that increasing the amount of coherences generated and detected gives a strategy to enhance the deviation of the quantum walk from the classical realm.

\section*{Discussion and conclusion}

Several strategies can be followed to identify distinctly quantum resources, i.e., those features of quantum systems which cannot be reproduced by any classical means.
Different facets of non-classicality can be in fact taken into account, such as the statistics of non-commuting observables \cite{Alicki2008,Facchi2013} or non-local configurations \cite{Bell1987}, the negativity of pseudo-probability distributions in phase space \cite{Glauber1963,Kenfack2004} or the establishment of quantum correlations in the course of the evolution \cite{Chen2018,Chen2019}.
Here, we consider sequential (projective) measurements of one and the same observable at different times, identifying non-classicality with the violation of the Kolmogorov consistency conditions. Essentially tracing back to the seminal works by Leggett and Garg \cite{Leggett1985,Leggett2002}, non-classicality is understood as measurement invasiveness, i.e., with the impossibility, \emph{even in principle}, to access the measured observable without altering the state of the system and hence the subsequent statistics. Such an approach avoids full tomographic procedures, as well as the detailed microscopic modeling of the system at hand, thus relying on a limited knowledge \emph{a-priori}.

Our main result consists in the theoretical and experimental demonstration of a quantitative connection -- \cref{eq:ck} -- between non-classicality, in the sense of the violation of the Kolmogorov conditions, and the amount of quantum coherence which is first generated and later turned into populations. This connection provides us with a general scheme to tune
the degree of non-classicality
in experimental settings via the manipulation of quantum coherence, as we illustrated explicitly in a time-multiplexing quantum walk. In particular, we achieved a full control 
over the generated and detected coherences and on the statistics involving intermediate measurements during the walk. 
This has been possible by virtue of the use of the electro-optic modulators to out-couple deterministically the walker, depending on its position and coin values, with extremely good extinction ratio. Besides its conceptual interest, our investigation proves the extremely high degree of control on our time-multiplexing setup, which allows us to access and manipulate multi-time non-classicality, as clearly shown by the data reported in Figs.\ref{Fig:Visualize} and \ref{Fig:KCPlot}.   

Importantly, our approach points out the relevant feature of quantum coherence which is linked to non-classicality when one considers sequential measurements of a reference observable \cite{Smirne2018}. Despite being related to properties emerged in the context of the resource theory of coherence \cite{Baumgratz2014,Streltsov2017}, the capability of the dynamical maps to generate coherences and turn them into populations, as quantified via ${\sf{C}}_{x_0,p}$ defined in \cref{eq:C0}, cannot be identified with any of such properties. In particular, dynamical maps for which ${\sf{C}}_{x_0,p}= 0$ include as special cases the maps which cannot generate coherences (i.e., which are maximally incoherent \cite{AAberg2014}), as well as those which cannot detect coherences \cite{Liu2017}. However, the notion of ''non-classical dynamics'', as defined by the condition ${\sf{C}}_{x_0,p}\neq 0$, emphasizes that a deviation from classicality can be detected 
only if first some coherences are generated and later \emph{the same coherences} are turned into populations.

Finally, we would like to emphasize that our analysis allows us to obtain stronger conclusions than the standard comparison between the classical and the quantum random walk of a single walker. The standard comparison is in fact based on the one-time position probability distribution of the walker (in our notation, on $\sum_c P_{x_0,p}(x,c,N)$) and it identifies classical and quantum walks as those where such probability distribution has, respectively, a Gaussian and a double-peaked shape \cite{Kempe2003}. However, it has been shown \cite{deFalco2008,Montero2017,Andrade2019} that a double-peaked distribution can be obtained also in a fully classical walk, if one takes into account a time inhomogenous distribution of the jump rates, i.e., in the presence of a time-inhomogeneous Markovian classical walk.
On the other hand, the analysis of the multi-time statistics we performed does discriminate between a genuinely quantum walk and 
any possible classical modeling of it, including the mentioned 
Markovian time-inhomogeneous models, or even any non-Markovian description, which introduces memory effects in the walk, as long as the Kolmogorov conditions are satisfied.

Our results will be hopefully a useful step toward the complete identification of the distinctly non-classical aspects of quantum physics. Future analysis will be focused on the investigation of more complex open-system evolutions, possibly involving quantum non-Markovian scenarios \cite{Milz2019,Strasberg2019}.

\ack{We thank Benjamin Brecht and Benjamin Desef for useful discussions.
The work at Ulm is supported by the Alexander von Humboldt Foundation and the ERC Synergy grant BioQ (No. 319130). 
D.E. is supported by the Swiss National Science Foundation SNSF (Grant No. P2SKP2\_184068). 
The Integrated Quantum Optics group in Paderborn acknowledges financial support from European Commission with the ERC project QuPoPCoRN (no.\ 725366) and from the Gottfried Wilhelm Leibniz-Preis (grant number SI1115/3-1).
}

\section*{References}

\providecommand{\newblock}{}

\appendix

\section{Quantitative relation between non-classicality and coherence
for open-system evolutions}
\label{App:Relation}
Here we show that \cref{eq:ck} 
can be derived under more general conditions than the discrete-time quantum walk on a line we considered in the experimental setup.

Let us take a system whose state at time $t$ is denoted as $\rho_t$, 
and whose evolution between 0 and $t$ is governed by the Lindblad equation \cite{Breuer2002,Rivas2012}
\begin{eqnarray}
\frac{d}{d t} \rho_t=\mathcal{L}\rho_t = -i \left[ \hat{H}, \rho_t \right] 
+\sum_j c_j\left(\hat{L}_j \rho_t \hat{L}^{\dag}_j -\frac{1}{2}\left\{\hat{L}^{\dag}_j \hat{L}_j, \rho_t\right\}\right),
\end{eqnarray}
where $ \hat{H}= \hat{H}^{\dag}$ and $\hat{L}_j $ are linear operators on the Hilbert space
associated with the system, and $c_j \geq 0 \,\, \forall j$; the corresponding evolution 
can be represented by
\begin{equation}
\rho_t=e^{\mathcal{L} t} \rho_0
\end{equation}
in terms of the generator $\mathcal{L}$ of the dynamics.
The discrete-time unitary case described
in the main text is obtained when
$c_j =0$ for any $j$, so that $e^{\mathcal{L} t}=\mathcal{U}_t$ 
and $\mathcal{U}_t\rho_0 = \hat{U}_t\rho_0\hat{U}^{\dag}_t$,
with $ \hat{U}_t=e^{-i \hat{H}t}$, and considering
$t= N \delta t$, for a fixed $\delta t$ and different values of $N$.

Moreover, we consider sequential projective measurements 
of the observable $X$, associated
with the non-degenerate self-adjoint operator $\hat{X}=\sum_x x \ketbra{x}{x}$.
For a fixed initial state $\rho_0$, which is diagonal in the eigenbasis of $\hat{X}$,
\begin{equation}\label{eq:is}
\rho_0 = \sum_x p_x  \ketbra{x}{x},
\end{equation}
we quantify the amount of coherences (with respect to eigenbasis of $\hat{X}$) which are generated 
up to a time $s$ and subsequently turned into populations at time $t$ via
the quantity
\begin{align}\label{eq:cgd}
{\sf{C}}_{\rho_0} = \left\|\left(\Delta \circ e^{\mathcal{L}(t-s)} \circ \Delta \circ  e^{\mathcal{L}s} 
-\Delta \circ e^{\mathcal{L}t}\right)\rho_0\right\|_1,
\end{align}
where $\Delta = \sum_x \ketbra{x}{x}\cdot \ketbra{x}{x}$
is the total dephasing map with respect to $\hat{X}$.
The difference between the two terms in the definition of ${\sf{C}}_{\rho_0}$
describes how the action of the total dephasing map 
$\Delta$ at an intermediate time $s$, which destroys the coherences generated
up to that time, will impact the population at a later time $t$.
Indeed, this provides us with a quantifier of those
coherences generated up to the time $s$ which are mapped into populations at $t$.
The situation we treated in the main text is recovered by identifying $x$ with both 
the position and the coin values of the walker, setting $t=N \delta t$
and $s = M\delta t$, as well as restricting $\rho_0$ to the state in \cref{eq:init}; 
further assuming a unitary evolution,
Eq.~(\ref{eq:cgd}) reduces in fact to \cref{eq:C0}.

Now, denote as $P_{\rho_0}(x,t)$ the probability
to get the outcome $x$ with a projective measurement of the observable $X$ at time $t$,
having the initial state $\rho_0$,
i.e.,
\begin{equation}
P_{\rho_0}(x,t) = \mbox{tr}\left\{\mathcal{P}_x e^{\mathcal{L}t} \rho_0\right\}
= \bra{x} e^{\mathcal{L}t} \rho_0 \ket{x},
\end{equation} 
where we introduced the projector super-operator 
$\mathcal{P}_x=\ketbra{x}{x} \cdot \ketbra{x}{x}$ and every super-operator
acts on everything at its own right.
One can then easily see that, analogously to \cref{eq:C},
the quantity in Eq.~(\ref{eq:cgd}) can be equivalently written as
\begin{align}\label{eq:CRho}
{\sf{C}}_{\rho_0} = 
 \sum_{x} \Big|  \sum_{y} 
P_{y}(x,t-s) P_{\rho_0}(y,s) 
 - P_{\rho_0}(x,t) \Big|,
\end{align}
where again with a slight abuse of notation we denote as $P_{x}$ the probabilities when the initial state is $\ket{x}$.

Let us move to the statistics associated with sequential measurements of $X$
at different times. Consider in particular the two-time joint probability distribution $P_{\rho_0}(x,t; y, s)$ of
having outcome $y$ at the intermediate time $s$ and $x$ at time $t\geq s$ for an initial state $\rho_0$,
as well as the conditional probability $P_{\rho_0}(x, t | y,s)$
of having outcome $x$ at time $t$,
given that there was the outcome $y$ at time $s$ (for an initial state $\rho_0$),
which is of course defined as 
$$
P_{\rho_0}(x, t | y,s) = \frac{P_{\rho_0}(x,t; y, s)}{P_{\rho_0}(y, s)}.
$$
We can then introduce the non-classicality quantifier
\begin{align}
{\sf{K}}_{\rho_0} &=  \sum_{x} \Big | \sum_{y} 
P_{\rho_0}(x, t ; y,s)
 - P_{\rho_0}(x,t) \Big| \nonumber\\
 &= \sum_{x} \Big |  \sum_{y} 
P_{\rho_0}(x, t | y,s) P_{\rho_0}(y,s) 
 - P_{\rho_0}(x,t) \Big|,\label{eq:ksupp}
\end{align} 
which indeed reduces to \cref{eq:K} for $\rho_0$
given by \cref{eq:init}, $t=N \delta t, s = M \delta t$ and $x$ denoting both the position and coin of the walker.

Now, if we assume that the joint probability can be written according to the quantum regression theorem,
as~\cite{Lax1968,Li2018}
\begin{equation}\label{eq:joint}
P_{\rho_0}(x,t; y, s) = 
\mbox{tr}\left\{\mathcal{P}_{x} e^{\mathcal{L}(t-s)}\mathcal{P}_{y} e^{\mathcal{L}s} \rho_0\right\},
\end{equation}
we directly get
\begin{equation}\label{eq:xx}
P_{\rho_0}(x, t | y,s) = P_y(x,t-s),
\end{equation}
which leads us to, see Eqs.\eqref{eq:CRho} and \eqref{eq:ksupp},
\begin{equation}
{\sf{K}}_{\rho_0} = {\sf{C}}_{\rho_0}.
\end{equation}
We thus conclude that \cref{eq:ck} can be properly generalised,
whenever we consider sequential projective measurements (at generic times $s$
and $t$)
of any non-degenerate observable of a system whose state is initially diagonal
(in the eigenbasis of the measured observable) and then undergoes a Lindblad evolution,
and the two-time probabilities satisfy the quantum regression theorem.
Therefore, the qualitative correspondence proven in~\cite{Smirne2018} for the same conditions, is promoted to a quantitative correspondence, which is also more tractable in experiments.

\section{Equivalence of Coherent Light and Single Photons} \label{sec:equivalence_one_ini_pos}
\label{Sec:AppendixEquivalence}

The objective of our experimental work is the investigation of the evolution of the wave function of a photonic walker, i.e., a single photon. 
Here we show that by investigating coherent pulses of indistinguishable photons in the same state $ \ket{\Psi}$, we observe the same evolution as for single photons as detailed in~\cite{Schreiber2014}.

We start by defining the creation operator $\hat{a}_i^\dagger$ which creates a photon in the $i$-th mode of the vacuum state $| 0\rangle$:
\begin{equation}
\hat{a}_i^\dagger \ket{\mathrm{vac}} = \ket{0_1, 0_2, ..., 1_i, 0_{i +1}, ..., 0_{2(N+1)}}.
\label{eq:creation}
\end{equation}
For an evolution over $N$ steps the number of possibly occupied position modes equals $N+1$.
Taking the polarisation into account, we consider a space $\mathcal{H} = \mathcal{H}_\text{x} \otimes \mathcal{H}_\text{c}$ with the dimension $2(N+1)$.

The evolution of a single photon is governed by a passive linear optical transformation, whose effect on one photon is independent of how many photons are evolving.
More precisely, consider an evolution of the photon that can be described with the unitary evolution operator $\hat{U}$
acting on $\mathcal{H}$
and the creation operator $\hat{a}_{0}^\dagger$ of the initial state:
after the $N$-th step, the evolved single-photon state is given by 
\begin{equation}
\hat{U}^N \hat{a}_{0}^\dagger | 0 \rangle = \sum_i A_i(N) \hat{a}_{i}^\dagger | 0 \rangle,
\label{eq:single_evolution}
\end{equation}
where
$A_i(N)$ denotes the probability amplitude of the $i$-th mode in step $N$. Accordingly, the probability $P(m, N)$ to measure the walker in mode $m$ in the $N$-th step, is given by the following expression:

\begin{align}
\begin{split}
P(m, N) =\,& | \langle 1_m | \sum_i A_i(N) \hat{a}_{i}^\dagger | 0 \rangle|^2 = | A_m(N)|^2.
\end{split}
\label{eq:single_probability}
\end{align}

In order to simulate the evolution of a single photon with coherent light, the presence of one photon must not influence the evolution of another. Thus, we take a look at the evolution of the wave function for $n$ photons, which is given by the following term:
\begin{equation}
\frac{1}{\sqrt{n !}} ( \hat{U}^N \hat{a}_{0}^\dagger)^n | 0 \rangle = \frac{1}{\sqrt{n !}} ( \sum_i A_i(N)\hat{a}_{i}^\dagger)^n | 0 \rangle.
\label{eq:p_evolution}
\end{equation}
To see whether the probability distribution for the outcome of the experiment is altered by additional photons, we determine the probability $P(m, N)$ of a measurement event in the $m$-th mode after $N$ steps for the simplest case of $n = 2$. 
We thus have
\begin{align}
P(m, N)
=\,& | \sum_{j \neq m} \langle 1_m, 1_j | \frac{1}{\sqrt{2}} ( \sum_{i} A_{i}(N) \hat{a}_i^\dagger )( \sum_{k} A_{k}(N) \hat{a}_{k}^\dagger ) | 0 \rangle |^2
\nonumber \\
&+ | \langle 2_m | \frac{1}{\sqrt{2}} ( \sum_{i} A_{i}(N) \hat{a}_i^\dagger )( \sum_{k} A_{k}(N) \hat{a}_{k}^\dagger ) | 0 \rangle |^2 \nonumber\\
=\,& \sum_{j \neq m} \left[| A_j(N) |^2 |A_m(N)|^2 \right]+ | A_m(N)|^4 \nonumber
\\
=\,& |A_m(N)|^2 \sum_{j} | A_j(N) |^2
=\, |A_m(N)|^2.
\label{eq:2_probability}
\end{align}
The above expression, derived for two photons, equals Eq.\eqref{eq:single_probability}, which describes the one-photon case. Consequently, we see that the probability for a measurement event in mode $m$ is unaffected by the presence of another photon. Knowing that an additional photon does not have an effect, the statement can be extended to arbitrarily large number of indistinguishable photons that are initially in the same mode of $| \Psi \rangle$.\\

In the next step, we examine the evolution of coherent states,
which, in the photon-number representation, reads (with $\alpha$ being the eigenvalue of the creation operator):

\begin{equation}
\begin{split}
| \alpha \rangle = e^{- |\alpha|^2 / 2} \cdot e^{\alpha \hat{a}_i^\dagger} | 0 \rangle = e^{- |\alpha|^2 / 2} \sum_{n = 0}^{\infty} \alpha^n \frac{\hat{a}_i^{\dagger n}}{n !} | 0 \rangle,
\end{split}
\label{eq:coherent_state}
\end{equation}
where, crucially, we consider the case where all photons are created in the same mode. 
The resulting quantum walk is indeed obtained by
including the single-photon evolution operator $\hat{U}^N$ as
\begin{align}
\begin{split}
e^{- |\alpha|^2 / 2} \cdot e^{\alpha \hat{U}^N \hat{a}_{0}^{\dagger}} | 0 \rangle 
=\,&e^{- |\alpha|^2 / 2} \cdot e^{\alpha \sum_i A_i(N) \hat{a}_{i}^{\dagger}} | 0 \rangle \\
=\,& e^{- |\alpha|^2 / 2} \sum_{n = 0}^{\infty} \frac{\alpha^n}{n !} ( \sum_i A_i(N) \hat{a}_{i}^{\dagger})^n | 0 \rangle.
\end{split}
\label{eq:coherent_evolution}
\end{align}
Each term in the final line of \cref{eq:coherent_evolution} is of the same form as \cref{eq:p_evolution},
which allows us to determine the probability of a measurement event independent of the presence of another photon. Consequently, we can determine $P(m, N)$ analogously to what done in \cref{eq:2_probability}, thus getting
\begin{align}
\begin{split}
P_{\alpha}(m, N) =\,& 
| \langle 1_m | e^{- |\alpha|^2 / 2} \cdot \alpha \sum_i A_i(N) \hat{a}_{i}^{\dagger} | 0 \rangle |^2
=\, e^{- |\alpha|^2} \cdot |\alpha|^2 |A_m(N)|^2.
\end{split}
\label{eq:coherent_probability}
\end{align}

Thus, the difference of a coherent state evolution as compared to a quantum walk conducted with single photons is merely a pre-factor depending on $\alpha$, which affects the overall probability of a measurement event, but not their distribution over the modes. The relation found here is crucial for our experimental work as it shows that quantum walks of single photons can be simulated with coherent light. Consequently, the experiment does not require a single photon source, saving a lot of experimental resources.
The results obtained for a single occupied input position of course do not mean that there is never a difference between a quantum walk conducted with coherent light and a quantum walk with single photons. 
As an example, when considering coincidences in a quantum walk initialised at more than one position, qualitative differences between coherent states and single photons might arise.
As our current experiment does not rely on coincidences, but on uncorrelated intensity measurements only, the mapping from coherent states to single photons is valid.
A more detailed discussion of using coherent states to simulate single-particle quantum walks can be found in~\cite{Schreiber2014}.

\section{Monte-Carlo-based Error Estimation} \label{sec:monte_carlo_error}
\label{App:Errors}

In order to obtain a numerical estimation for the effect of experimental inaccuracies, we conduct a Monte-Carlo-based error estimation. 
The error estimation procedure is based on the assumption that the main source of errors in the experiment is the imprecision in the quantum walk parameters, including the coin angle and losses at out-coupling.
This assumption is reasonable for our optical setup because the variance from shot noise and detection imperfection is much smaller than the above-mentioned imprecision.
The quantum-walk parameters are subject to measurement imprecision because of how they are set or measured: the coin angle is set manually using a scale with finite accuracy, while the losses from coupling inefficiencies are determined with a power-meter exhibiting an uncertainty as well.
In addition, the opto-mechanical components in the setup might show a slight drift during the intervals in which the measurements are taken. 
Finally, the uncertainty on the transmission of the position-dependent out-coupling originates from the fact that the extinction ratio of the switchings conducted by the electro-optic modulators can only be determined up to some error.

Note that it is enough to consider only mechanisms that introduce inhomogeneous losses that are either dependent on a certain coupling or a certain position, because any homogeneous losses (those acting uniformly on all optical modes) will not affect the normalized probability distributions that we use here.
Referring to the analysis performed in \ref{Sec:AppendixEquivalence}, inhomogeneous losses alter the probability amplitudes $A_i(N)$ of the individual modes in Eq.~(\ref{eq:coherent_probability}), whereas homogeneous losses merely alter the global pre-factor, but not the individual amplitudes.

To perform the error estimation, we simulate the evolution of the walker according to the quantum walk evolution operator and generate for each configuration multiple, in this case 1000, different instances with varying parameters chosen randomly within a defined range of uncertainty. 
For the angle of the coin we assume an inaccuracy of $0.5^{\circ}$, for the coupling efficiencies between different modes an uncertainty of $2 \%$ and also $2 \%$ for the residual transmission of the position-dependent out-coupling.
Once this myriad of evolutions based on sightly different walk parameters is simulated, we calculate the values of the coherence and non-classicality quantifiers for each simulated evolution. 
Thus, we obtain a distribution of the quantifiers with respect to different parameters.
The standard deviation of this distribution is the Monte-Carlo estimate of the uncertainty that we plot
as vertical and horizontal error bars in \cref{Fig:KCPlot}.






\section{Explanation of imperfect matching observed in Fig.~\ref{Fig:KCPlot}} \label{App:KDisplacement}

\begin{figure}[htb]
\centering
	\includegraphics[width=0.75\columnwidth]{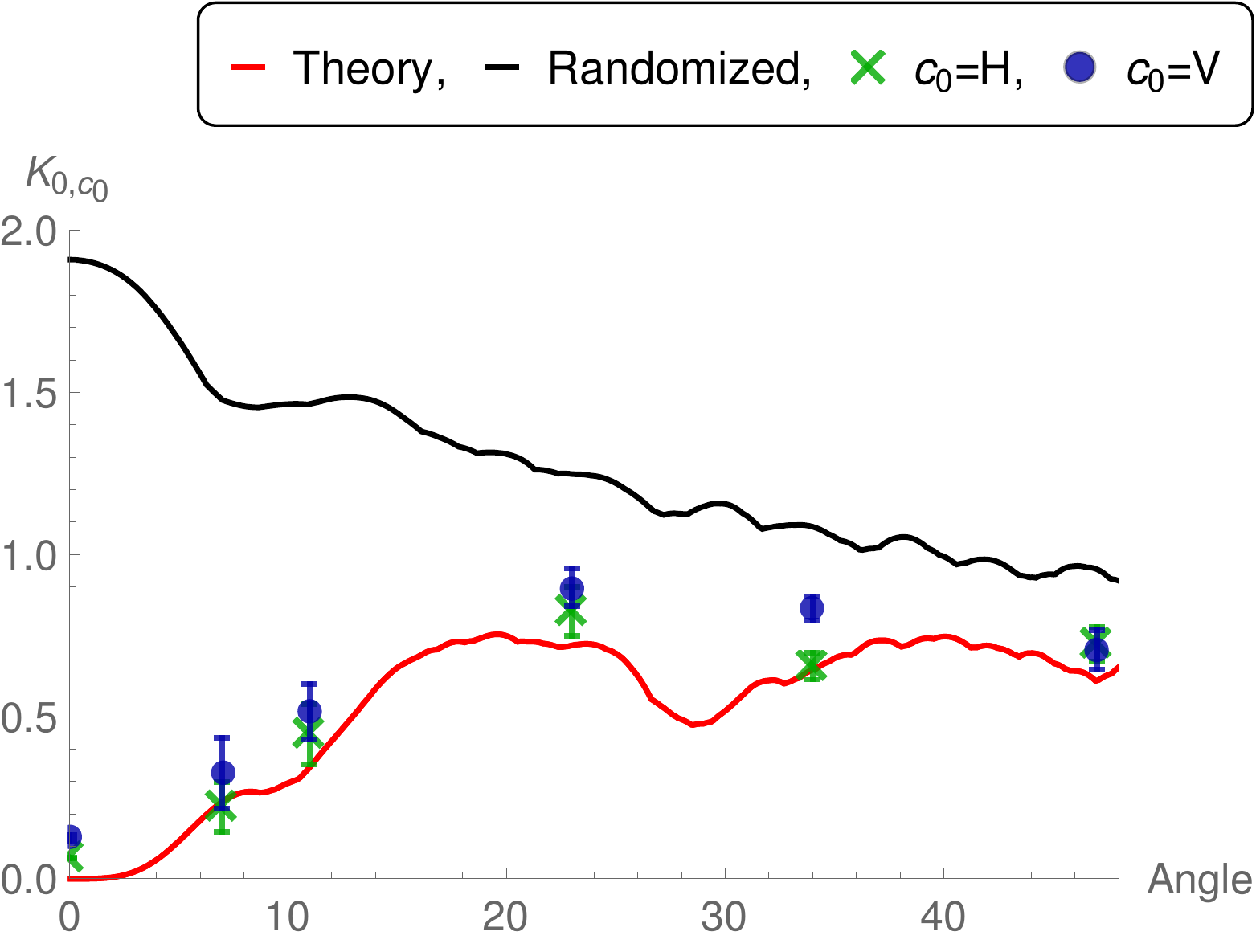}
	\caption{
	\textbf{The quantifier ${\sf{K}}_{0,c_0}$ plotted against the angle of the coin.}
	We plot on the ordinate the quantifier ${\sf{K}}_{0,c_0}$ of coherences generated and detected 
	as defined in \cref{eq:K} and on the abscissa the angle of the coin used, for initial polarisation $c_0=H$ (circles) and $c_0=V$ (crosses);
	the symbols correspond to the experimental data.
	The red line represents the theoretical prediction for the different angles \cref{eq:ck} and the black line represents the theoretical values obtained by assuming a randomizing intermediate measurement.
	Error bars are generated from the Monte-Carlo approach detailed in \ref{App:Errors}.
	A slight systematic deviation observed in the higher-than-expected ${\sf{K}}_{0,c_0}$ values can be explained by imperfect intermediate measurements as described in this section.
	}
	\label{Fig:KPlot}
\end{figure}

In Fig.~\ref{Fig:KCPlot}, even though the data confirms the theoretical prediction within the error bars in the regime of small coin angles, a displacement of the experimental values relative to the theoretical prediction is apparent. 
Here we explore the possible causes of this imperfect matching between theory and experiment. 

\begin{figure}[htb]
\centering
	\includegraphics[width=0.75\columnwidth]{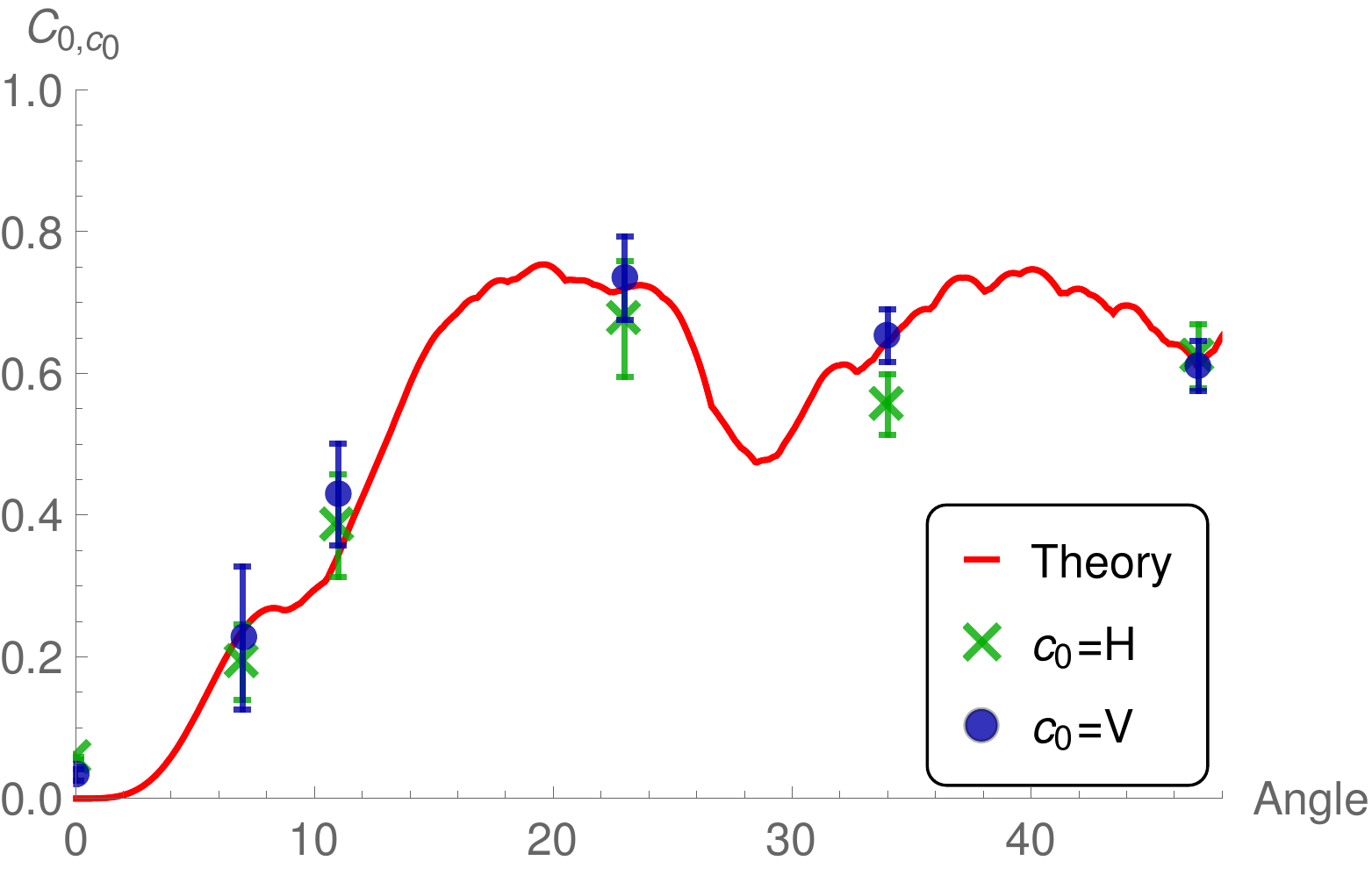}
	\caption{
	\textbf{The quantifier ${\sf{C}}_{0,c_0}$ plotted against the angle of the coin.}
	We plot on the ordinate the quantifier ${\sf{C}}_{0,c_0}$ of coherences generated and detected 
	as defined in \cref{eq:C} and on the abscissa the angle of the coin used, for initial polarisation $c_0=H$ (circles) and $c_0=V$ (crosses);	
	the symbols correspond to the experimental data.
	The red line represents the theoretical prediction for the different angles \cref{eq:ck}.
	Error bars are generated from the Monte-Carlo approach detailed in \ref{App:Errors}.
	}
	\label{Fig:CPlot}
\end{figure}

As this error is more pronounced for the Kolmogorov values than for the coherence measure, it could result from imperfect intermediate measurements as these would effect only the Kolmogorov and not the coherence values. 
A perfect projective measurement in the context of our quantum walk experiment is one that would couple out all the light from all the modes except one.
However, in real experiments, this extinction is often imperfect as a small fraction of the light continues to propagate in the out-coupled modes.

In more detail, recall that 
${\sf{K}}_{0,c_0}$
measures the difference between the statistics after an unperturbed evolution and the statistics that comes from an  evolution having been measured at some intermediate time. 
If this measurement perturbs the evolution, 
${\sf{K}}_{0,c_0}$
is not zero. 
Such a perturbation can stem from quantum-mechanical effects as we expect in an ideal experiment, but this perturbation can also see a contribution from an imperfect measurement. 

Focusing on imperfect measurements, consider hypothetically a measurement that projects the state into one that is completely random.
In our case, comparing the highly structured probability distributions we get from the unperturbed evolution with evolving a flat distribution with values of $1/[2(N+1)]=1/22$ after the intermediate measurement, we get the values displayed in Table~\ref{tab:kolm_shitf} and visualised in Fig.~\ref{Fig:KPlot}. of ${\sf{K}}_{0,c_0}$ for the different angles and initial polarizations.
More sophisticated models of imperfect measurement can be considered but the current simple model already provides some qualitative understanding.

\begin{table}[h!]
    \centering
\begin{tabular}{|l r|c|c l r|c|}    \hline   
        $\theta$&$c_0$&Theory&Experiment& &Error&Randomizing \\
        \hline
        $0^\circ$&
        V & 
        0.000 &
        0.064& 
        $\pm$ & 
        0.001& 
        1.909  \\
        \hline
        $0^\circ$&
        H & 
        0.000 &
        0.127 & 
        $\pm$ & 
        0.009& 
        1.909  \\
        \hline
        $7^\circ$&
        V & 
        0.237 &
        0.221& 
        $\pm$ & 
        0.076& 
        1.477   \\
        \hline
        $7^\circ$&
        H & 
        0.237 &
        0.325& 
        $\pm$ & 
        0.110& 
        1.477   \\
        \hline
        $11^\circ$&
        V & 
        0.343 &
        0.446& 
        $\pm$ & 
        0.093& 
        1.464   \\
        \hline
        $11^\circ$&
        H & 
        0.343 &
        0.514& 
        $\pm$ & 
        0.086& 
        1.464   \\
        \hline
        $23^\circ$&
        V & 
        0.720 &
        0.825& 
        $\pm$ & 
        0.075& 
        1.248  \\
        \hline
        $23^\circ$&
        H & 
        0.720 &
        0.8298& 
        $\pm$ & 
        0.059& 
        1.248  \\
        \hline
        $34^\circ$&
        V & 
        0.644 &
        0.655& 
        $\pm$ & 
        0.041& 
        1.085  \\
        \hline
        $34^\circ$&
        H & 
        0.644 &
        0.833& 
        $\pm$ & 
        0.038& 
        1.085  \\
        \hline
        $47^\circ$&
        V & 
        0.612 &
        0.724& 
        $\pm$ & 
        0.053& 
        0.954  \\
        \hline
        $47^\circ$&
        H & 
        0.612 &
        0.705& 
        $\pm$ & 
        0.061& 
        0.954  \\
        \hline
    \end{tabular}
\caption{Values for ${\sf{K}}_{0,c_0}$, for different coin angles and initial polarizations, from theoretical prediction, experimental results and theoretical prediction assuming a randomizing intermediate measurement.}
    \label{tab:kolm_shitf}
\end{table}

We see that the values of ${\sf{K}}_{0,c_0}$  for a randomizing intermediate measurement are consistently higher than the theoretical predictions. 
If we now consider an intermediate measurement that sees a small imperfection, then the value of 
${\sf{K}}_{0,c_0}$ 
is likely to be shifted towards the value one would get introducing a totally random state.
In other words, assuming that the out-coupling at the intermediate time is not perfect, one has a further perturbation due to the measurement and it is not surprising to have a value for
${\sf{K}}_{0,c_0}$,
which is slightly shifted towards the random case and is thus consistently higher then the theoretical prediction.

Note that such a perturbation due to the measurement does not affect the measure 
${\sf{C}}_{0,c_0}$,
as this is calculated using the unperturbed evolutions only; and indeed, ${\sf{C}}_{0,c_0}$ is remarkably close to the theoretical prediction (see Fig.~\ref{Fig:CPlot}).

\end{document}